\documentclass[preprint]{aastex}
\begin{document}
\title{Vertical Structure of Neutrino Dominated Accretion Disks\\
and Neutrino Transport in the disks}
\author{Zhen Pan and Ye-Fei Yuan\thanks{yfyuan@ustc.edu.cn}}
%\email{yfyuan@ustc.edu.cn}
%\email{panzhen@mail.ustc.edu.cn}
\affil{Key Laboratory for Research in Galaxies and Cosmology CAS, \\
Department of Astronomy, University of Science and Technology of China,\\
 Hefei, Anhui 230026, China}
\begin{abstract}
We investigate the vertical structure of neutrino dominated accretion disks by self-consistently considering
the detailed microphysics, such as
the neutrino transport, vertical hydrostatic equilibrium, the conservation of lepton number,
as well as the balance between neutrino cooling, advection cooling and viscosity heating.
After obtaining the emitting spectra of neutrinos and antineutrinos by solving
the one dimensional Boltzmann equation of neutrino and antineutrino transport in the disk,
we calculate the neutrino/antineutrino luminosity and their annihilation luminosity. We find that the total neutrino and antineutrino luminosity is about $10^{54}$ ergs/s and their annihilation luminosity is about $5\times10^{51}$ ergs/s with an extreme accretion rate $10 M_{\rm {sun}}$/s and an alpha viscosity $\alpha=0.1$.
In addition, we find that the annihilation luminosity is sensitive to the accretion rate and will not exceed $10^{50}$ ergs/s 
which is not sufficient to power the most fireball of GRBs, if the accretion rate is lower than $1 M_{\rm {sun}}$/s.
Therefore, the effects of the spin of black hole or/and the magnetic field in the 
accretion flow might be introduced to power the central engine of GRBs.

\end{abstract}
\maketitle

\section{Introduction}
Gamma ray burst(GRBs) are extremely high energy releasing phenomena in the universe and are usually divided into two classes \citep{1993ApJ...413L.101K,2004IJMPA..19.2385Z,2004RvMP...76.1143P,2007PhR...442..166N}:
short GRBs ($T_{90}<2$s) and long GRBs ($T_{90}>2$s).
Numerous models have been proposed to explore the central engines of GRBs,
and one of the mostly discussed model is the neutrino dominated accretion flows (NDAFs) with a hyper accreting stellar massive black hole
with accretion rate $0.1\sim10 M_{\rm {sun}}$/s. Due to the high density and high temperature in the inner part of NDAFs,
the optical depth of photons is very large and photons are completely trapped,
then neutrinos and antineutrinos become the most promising candidates that carry away thermal energy and cool the disk.
The annihilation of neutrino pairs above the disk is believed to be the energy source of GRBs.
\citet{1992ApJ...395L..83N} first proposed that neutrino pairs annihilation into electron pairs
during the merger of compact objects binaries may power GRBs.
After that, \citet{1999ApJ...518..356P} investigated NDAFs under the assumption that the disk is transparent to neutrinos,
but they also pointed out that the assumption fails when the accretion rate is higher than $1 M_{\rm {sun}}$/s .
\citet{2002ApJ...579..706D} improved the model by using a simplified neutrino transport model
which was believed to bridge the neutrino optically thin limit and optically thick limit.
\citet{2007ApJ...657..383C} improved the model further by dealing with the neutrino emission and chemical composition
in the optically thin regime, optically thick regime and intermediate regime separately.
Though many works on NDAFs have confirmed the validity of the NDAFs model as the central engine of GRBs
\citep{2001ApJ...557..949N,2002ApJ...577..311K,2005ApJ...632..421L,2007ApJ...664.1011J,2006ApJ...643L..87G,2007ApJ...661.1025L,2008ApJ...676..545L,2010ApJ...709..851L,2010ApJ...718..841Z,2009ApJ...703..461Z},
there are still some uncertainties including: 1, The distribution of electron fraction.
In many previous works, which is assumed to be a constant value, for example, 0.5, throughout the disk.
While the electron fraction has a great effect on the emission of neutrinos and antineutrinos,
as shown by \citet{2002ApJ...577..311K,2005ApJ...629..341K,2007ApJ...661.1025L}, so need more cautious disposal.
2, Neutrino transport and neutrino spectra. The most commonly used approximation was the simplified neutrino transport model introduced in
\citet{2002ApJ...579..706D}.
In their model, the difference between the neutrino and antineutrino transport in the disk was neglected, but as shown in \cite{2012PhRvD..85f4004P},
the precise spectra of neutrino and antineutrino sensitively determines the annihilation luminosity of neutrino pairs.
3, The annihilation of neutrino pairs. The most common method for calculating the annihilation luminosity was originally introduced by \citet{1997A&A...319..122R}
 to calculate the annihilation luminosity during the merger of neutron star binaries.
 This method was applied in the calculation of the annihilation luminosity above NDAFs under the assumption
that the emission of neutrinos and antineutrinos are isotropic and symmetric \citep{1999ApJ...518..356P}.

It is evident that the most strict approach to determine the neutrino/antineutrino luminosity
and their annihilation luminosity above NDAFs is to build the two-dimensional disk model in which neutrino transport, vertical structure,
chemical evolution, thermal evolution and the distribution of electron fraction, mass density, and temperature are self consistently considered.

\citet{2007Ap&SS.311..185R} first investigated the vertical structure of NDAFs by using the Eddington approximation to deal with the neutrino transport
in the vertical direction, neglecting the contribution of advection term to the disk cooling,
and dealing with neutrino emission and chemical composition following the
similar method of \citet{2007ApJ...664.1011J} and \citet{2007ApJ...657..383C}.

Recently, \citet{2010ApJ...709..851L} also investigated the vertical structure of NDAFs with many simplifications on neutrino transport, equation of state and the annihilation efficiency.
The first one is that the neutrino emission is directly integrated to calculate the neutrino luminosity by neglecting the absorption of neutrinos,
which is viable in the neutrino optically thin limit \citep{1999ApJ...518..356P};
the second one, that a simplified equation of state $p=K\rho^{4/3}$ is used in the vertical direction,
which is viable when relativistic degenerate electrons dominate the pressure of the disk;
the third one that a toy annihilation efficiency of neutrino pairs $\eta\equiv L_{\nu\bar\nu}/L_{\nu}\propto V_{\rm ann}^{-1}$ is applied,
where $L_{\nu\bar\nu},L_{\nu}$ is the annihilation luminosity and neutrino luminosity before annihilation respectively,
and $V_{\rm ann}$ is the so called annihilation volume. With all the above simplifications,
the annihilation efficiency $\eta\equiv L_{\nu\bar\nu}/L_{\nu}$ can even reach $100\%$!
Obviously, the unrealistic result is caused by too many unrealistic assumptions.

In this work, we investigate the vertical structure of NDAFs, neutrino/antineutrino luminosity and their annihilation luminosity
by self-consistently considering the neutrino/antineutrino transport,
the vertical hydrostatic equilibrium, the precise equation of state, the chemical equilibrium
and the thermal balance between neutrino cooling, advection cooling and the viscosity heating
under the self similar assumption of the distribution of mass density and internal energy density in the radial direction
\citep{1973A&A....24..337S,1994ApJ...428L..13N,1995ApJ...444..231N}.
Especially, we strictly solve the Bolzmann equation to
deal with the neutrino transport, instead of taking
the assumption of the gray body spectra \citep{2007ApJ...664.1011J}
or the Eddington approximation \citep{2007Ap&SS.311..185R}.
Correspondingly, we can precisely obtain the energy spectra of neutrino pairs.
Combining the conservation of the number of lepton,
the distribution of chemical compositions are self-consistently and
accurately determined \citep{2007ApJ...657..383C,2007Ap&SS.311..185R}. 

This paper is organized as follows.
In \S II, we introduce the basic equations in our calculation,
including the Boltzmann equation
of neutrino/antineutrino transport, angular momentum equation, hydrostatic equilibrium equation,
 equation of state, lepton number conservation equation and thermal evolution equation.
In \S III, we briefly introduce our numerical methods to find the steady solution of the structure of the disk.
In \S IV, we list our numerical results of the disk structure, neutrino/antineutrino luminosity and their annihilation luminosity.
Conclusions and discussions are summarized in \S V.

\section{Basic Equations}
We assume a steady accretion disk with accretion rate $\dot M=10,1,0.1 M_{\rm{sun}}$/s  around a central black hole with mass $M=3.3M_{\rm{sun}}$
and adopt the standard $\alpha$ viscosity prescription of \citet{1973A&A....24..337S} with $\alpha=0.1$ for the viscous stress of the disk.
We discuss the structure of the disk and neutrino transport in cylindrical coordinate $(r,z,\phi)$
and we assume the inner boundary of the disk to be $r_{\rm in}=6 M$ and outer boundary to be $r_{\rm out}=100 M$.

\subsection{Boltzmann equation}
We solve the one dimensional Boltzmann equation of neutrino and antineutrino transport in the vertical direction
of the disk and obtain the energy dependent and direction dependent neutrino spectrum.
We define $f_+(z,p,\mu)$ and $f_-(z,p,\mu)$ to be the distribution function for up moving neutrinos/antineutrinos and down moving ones
respectively, where $z$ is the vertical coordinate of the disk, $p$ is the energy of neutrinos/antineutrinos, and $\mu=\cos(\theta)$
for up moving neutrinos/antineutrinos and $\mu=-\cos(\theta)$ for down moving ones, where $\theta$ is the angle of neutrinos/antineutrinos moving direction
to the vertical direction of the disk.
For the up-moving neutrinos/antineutrinos,
their distribution function is determined by \citep{2003PhRvD..68f3001S,2006NuPhA.777..356B,1982ApJS...50...23S}:
\begin{equation}
\mu \frac{\partial f_+(z,p,\mu)}{\partial z}=\lambda_a \left[f^{\rm{eq}}(T(z),\mu_{\rm{eq}},p)-f_+(z,p,\mu)\right]+\lambda_s\left[-f_+(z,p,\mu)+\frac{1}{2} \int_0^1 d\mu f_-(z,p,\mu)+f_+(z,p,\mu)\right],
\end{equation}
and for the down-moving ones, their distribution function is determined by
\begin{equation}
\mu \frac{\partial f_-(z,p,\mu)}{\partial z}=-\lambda_a \left[f^{\rm{eq}}(T(z),\mu_{\rm{eq}},p)-f_-(z,p,\mu)\right]-\lambda_s\left[-f_-(z,p,\mu)+\frac{1}{2} \int_0^1 d\mu f_-(z,p,\mu)+f_+(z,p,\mu)\right].
\end{equation}

Where $\lambda_a$ is the absorption coefficient and $\lambda_s$ is the scattering coefficient of neutrinos/antineutrinos,
and here $f^{\rm{eq}}=1/(\exp{((p-\mu_{\rm{eq}})/{kT})}+1)$ for neutrinos,
$f^{\rm{eq}}=1/(\exp{((p+\mu_{\rm{eq}})/{kT})}+1)$ for anti-neutrinos, where
$\mu_{\rm{eq}}=\mu_{\rm_{e}} + \mu_{\rm{p}} - \mu_{\rm{n}}$,  and $\mu_{\rm{e}},\mu_{\rm{p}},\mu_{\rm_{n}}$
is the chemical potential of electron, proton and neutron, respectively.

Because Urca process $\nu_e + n \leftrightarrow e^- + p$ and $\overline{\nu}_e + p \leftrightarrow e^+ + n$
dominate the creation and the absorption of neutrinos and antineutrinos,
and the neutrino/antineutrino scattering by neutrons and protons $(\nu_e,\overline{\nu}_e)+p\rightarrow(\nu_e,\overline{\nu}_e)+p$
and $(\nu_e,\overline{\nu}_e)+n\rightarrow(\nu_e,\overline{\nu}_e)+n$
dominates the scattering opacity \citep{1999ApJ...518..356P,2007ApJ...664.1011J,2007ApJ...661.1025L},
so we include no other neutrino/antineutrino processes.
Thus, for simplicity, in this paper we use $\nu$ and $\nu_e$, $\bar\nu$ and $\bar\nu_e$ interchangeably and it will not cause any confusion.
As for the explicit expression of the absorption coefficient $\lambda_a$
and scattering coefficient $\lambda_s$ of neutrinos/antineutrinos, please refer to \cite{2012PhRvD..85f4004P}.

Considering that the disk is symmetric about the equator plane, the boundary conditions for the distribution function $f(z,p,\mu)$ of neutrinos/antineutrinos
 can be written as $f_+(0,p,\mu)=f_-(0,p,\mu)$
and $f_-(H,p,\mu)=0 $, where $H$ is the upper boundary of the disk.

\subsection{Angular momentum equation and Vertical hydrostatic equilibrium equation }
Adopting $\alpha$ prescription, the tangential stress and angular momentum equation can be written as \citep{1973A&A....24..337S},
\begin{equation}
w_{r\phi}=\alpha\rho c_s^2,
\end{equation}

and

\begin{equation}
\rho\frac{d(\Omega r^2)}{dt}=\rho v_r\frac{d(\Omega r^2)}{dr}=\frac{1}{r}\frac{d(w_{r\phi} r^2)}{dr},
\end{equation}
where $w_{r\phi}$ is the viscous stress, $\rho$ is the mass density of the disk, $c_{\rm{s}}$ is the acoustic speed, $v_r$ is the radial drift velocity and $\Omega$ is Kepler angular velocity at radius $r$.
Integrating Eq.(4) over radius $r$ and height $z$, we obtain
\begin{equation}
2\pi\int_{-H}^{H}\int_{r_{\rm in}}^r r\rho v_r \frac{d(\Omega r^2)}{dr} dzdr=2\pi\int_{-H}^{H}\int_{r_{\rm in}}^r\frac{d(w_{r\phi} r^2)}{dr}dzdr,
\end{equation}
and by taking into consideration the steady accretion condition
\begin{equation}
\dot{M}=2\pi r\int_{-H}^{H}\rho v_rdz=\rm{const},
\end{equation}
and using torsion condition in the inner boundary $w_{r\phi}(r_{\rm in})=0$, the angular momentum equation Eq.(5) is transformed to be
\begin{equation}
\dot{M}\left(\Omega r^2-(\Omega r^2)_{\rm{in}} \right)=2\pi r^2\int_{-H}^{H} \alpha\rho c_{\rm{s}}^2 dz.
\end{equation}

The vertical hydrostatic equilibrium equation is simply as follows,
\begin{equation}
\frac{\rho GM}{r^2}\frac{z}{r}=-\frac{dp}{dz}.
\end{equation}

\subsection{Equation of state (EOS)}
In our calculation, the EOS of accreted gas including protons, neutrons, and electron pairs are determined by the exact Fermi-Dirac integral,
the EOS of neutrinos pairs are determined by the numerical integration of their distribution functions $[f_{+,-}(z,p,\mu)]_{\nu,\bar\nu}$,
and the EOS of photons is simply according to Eq.(15) and we do not include other kinds of particles in this work, especially, helium, which was included in
the most of the previous works (we will check the validity of neglecting the contribution of helium in \S IV):
\begin{eqnarray}
 p(\rho,Y_{\rm e},T)=p_{\rm n}+p_{\rm p}+p_{\rm e}+p_{{\rm e}^+}+p_{\rm {rad}}+p_{\nu}+p_{\bar\nu},\\
 u(\rho,Y_{\rm e},T)=u_{\rm n}+u_{\rm p}+u_{\rm e}+u_{{\rm e}^+}+u_{\rm {rad}}+u_{\nu}+u_{\bar\nu},
\end{eqnarray}
where $p$ and $u$ is the total pressure and total internal energy density respectively,
 $Y_e$ is the electron fraction $Y_e\equiv(n_{\rm e}-n_{\rm e^+})/(n_{\rm p}+n_{\rm n})$, and $T$ is the local temperature of the disk.

Specifically, the EOS of gas are expressed as \citep{2007ApJ...664.1011J},
\begin{eqnarray}
p_i=\frac{2\sqrt 2}{3\pi^2}\frac{(m_ic^2)^4}{(\hbar c)^3}\beta_i^{5/2}\left[F_{3/2}(\eta_i,\beta_i)+\frac{1}{2}\beta_i F_{5/2}(\eta_i,\beta_i) \right],\\
u_i=\frac{2\sqrt 2}{3\pi^2}\frac{(m_ic^2)^4}{(\hbar c)^3}\beta_i^{5/2}\left[F_{3/2}(\eta_i,\beta_i)+\beta_i F_{5/2}(\eta_i,\beta_i) \right],\\
n_i=\frac{\sqrt 2}{\pi^2}\left(\frac{m_ic^2}{\hbar c}\right)^3\beta_i^{3/2}\left[F_{1/2}(\eta_i,\beta_i)+\beta_i F_{3/2}(\eta_i,\beta_i) \right],
\end{eqnarray}
where $p_i,u_i,n_i$ is the pressure, internal energy density and number density of particle $i$ respectively ($i=\rm n,p,e,e^+$), $F_k$ is the Fermi-Dirac integral of order $k$, $\eta_i$ is the degeneracy parameter of particle $i$ ($\eta_i\equiv\mu_i^N/kT$, where $\mu_i^N$ is the chemical potential of particle $i$ not including the rest mass) and $\beta_i$ is the relativity parameter of particle $i$ ($\beta_i\equiv kT/m_ic^2$).

And the EOS of neutrino pairs and radiation are expressed as
\begin{eqnarray}
p_j=\frac{u_j}{3},\\
u_{\rm rad}=\frac{\pi^2}{15}\frac{(k T)^4}{(\hbar c)^3},\\
u_{\nu,\bar\nu}=\frac{2\pi}{h^3}\int\!\int{p^3(f_++f_-)_{\nu,\bar\nu}}dpd\mu,\\
n_{\nu,\bar\nu}=\frac{2\pi}{h^3}\int\!\int{p^2(f_++f_-)_{\nu,\bar\nu}}dpd\mu.
\end{eqnarray}
where $p$ is the energy of neutrinos and antineutrinos and $p_j$ is the pressure of particle $j$ ($j\equiv\rm rad,\nu,\bar\nu$).

\subsection{Lepton number conservation}
It is easy to write down the equation of lepton number conservation of fluid using the Eulerian description:
\begin{equation}
\frac{\partial {(n_{\rm b}Y_{\rm{lep}}})}{\partial t}+\mathbf v\cdot\nabla(n_{\rm b}Y_{\rm{lep}})+ \frac{\partial F_{\rm{lep}}}{\partial z}=0,
\end{equation}
where $n_{\rm b}$ is the number density of baryon, $Y_{\rm{lep}}$ is the fraction of lepton number
\begin{equation}
Y_{\rm{lep}}=\frac{n_{\rm e}-n_{\rm {e^+}}+n_{\nu}-n_{\overline{\nu}}}{n_{\rm b}},
\end{equation}
and $F_{\rm {lep}}$ is the lepton number flux in the vertical direction of the disk ( here $F_{\rm {lep}}$ is contributed by neutrinos and antineutrinos $F_{\rm {lep}}=F_{\nu}+F_{\bar\nu}$)
\begin{eqnarray}
F_{\nu}=\frac{2\pi c}{h^3}\int\!\int p^2(f_+-f_-)_{,\nu}\mu dpd\mu,\\
F_{\bar\nu}=-\frac{2\pi c}{h^3}\int\!\int p^2(f_+-f_-)_{,\bar\nu}\mu dpd\mu.
\end{eqnarray}

Due to the fact that the time scale of accretion is much longer than that of chemical evolution under the condition of the inner part
of NDAFs: $\rho\approx10^{10}\sim10^{11}$ g/cm$^3$, or $n_{\rm b}\approx10^{-5}\sim10^{-4}$ fm$^{-3}$, and $T\approx5\times10^{10}$K,
the time scale for the typical Urca process $p+e^-\rightarrow n+\nu_e$ is about $0.1\sim1$ ms ( see the Fig.2 of \citet{2005PhRvD..72a3007Y} )
which is much shorter than the time scale of accretion, so it is reasonable to neglect the advection term in the equation of lepton number conservation,
hence Eq.(18) is simplified to be
\begin{equation}
\frac{\partial {(n_{\rm b}Y_{\rm{lep}}})}{\partial t}+ \frac{\partial F_{\rm{lep}}}{\partial z}=0.
\end{equation}

\subsection{Thermal evolution equation}
The energy equation of fluid is written as
\begin{equation}
\frac{\partial u}{\partial t}+\mathbf v\cdot\nabla u=q_+-q_--p\nabla\cdot \mathbf v,
\end{equation}
or in the equivalent form
\begin{equation}
\frac{\partial u}{\partial t}=q_+-q_--q_{\rm adv},
\end{equation}
where $q_{\rm adv}=p\nabla\cdot\mathbf v+\mathbf v\cdot\nabla u$ is the advection cooling term and $q_+$ is the alpha-viscosity heating rate
\begin{equation}
q_+=\alpha\rho c_s^2 \left( r \frac{d\Omega}{dr}\right),
\end{equation}
and $q_-$ is neutrino cooling rate contributed by the energy flux of neutrinos and antineutrinos: $q_-=q_{\nu}+q_{\bar\nu}$
\begin{equation}
q_{\nu,\bar\nu}=\frac{2\pi c}{h^3}\frac{d}{dz}\int\!\int p^3(f_+-f_-)_{\nu,\bar\nu}\mu dpd\mu.
\end{equation}

We assume the velocity distribution to be
$v_r=-\alpha c_s^2/r\Omega$, $v_z=0$, and $\Omega=\sqrt{GM/r^3}$,
which is similar to the self similar radial distribution of the mass density of the gas pressure dominated thin disk \citep{1973A&A....24..337S},
we also assume the distribution of mass density and energy density to be self similar
$\rho\sim r^{-3/2}$ and $u\sim r^{-2}$ in the radial direction.
According to the mass conservation equation of steady flows $\rho\nabla\cdot\mathbf v+\mathbf v\cdot\nabla \rho=0 $,
we obtain $\nabla\cdot\mathbf v=3/2 (v_r/r)$, so the advection term $q_{\rm adv}$ is simplified to be
\begin{equation}
q_{\rm adv}=\frac{v_r}{r}\left(\frac{3}{2}p-2u\right),
\end{equation}
and we will check the validity of the self similar assumption in \S IV.

For simplicity, we adopt the value of acoustic speed at $z=0$ when calculating viscosity heating rate and radial velocity at any location,
i.e. more numerically economic expressions  $w_{r\phi}=\alpha\rho \left[c_s^2(z=0)\right]$ for the viscous stress
and $v_r=\alpha [c_s^2(z=0)]/r\Omega$ for the radial velocity,
and $c_s^2$ is the square of adiabatic sound speed, which is given by \citep{2003LRR.....6....4F}
\begin{equation}
c_s^2=\left[\frac{\partial p}{\partial (\rho+u)}\right]_{\rm ad}=\frac{1}{\rho+p+u}\left[\rho\left(\frac{\partial{p}}{\partial{\rho}}\right)_u+(p+u)\left(\frac{\partial{p}}{\partial{u}}\right)_{\rho}\right].
\end{equation}

\section{Numerical Methods}
\subsection{Two stream approximation}
The two-stream approximation is a simplification to the full Boltzmann equation that replaces the full direction dependent distribution
$f_+(z,p,\mu)$ and $f_-(z,p,\mu)$ by two streams with angle $\cos(\theta)=\pm 1/\sqrt3$ to the vertical direction.
Under the two stream approximation, the full Boltzmann equation is simplified to be
\begin{eqnarray}
\frac{1}{\sqrt3}\frac{\partial f_+(z,p)}{\partial z}=\lambda_a[f^{\rm{eq}}(T(z),\mu_{\rm{eq}},p)-f_+(z,p)]+\frac{1}{2}\lambda_s[f_-(z,p)-f_+(z,p)], \\
\frac{1}{\sqrt3}\frac{\partial f_-(z,p)}{\partial z}=-\lambda_a[f^{\rm{eq}}(T(z),\mu_{\rm{eq}},p)-f_-(z,p)]+\frac{1}{2}\lambda_s[f_-(z,p)-f_+(z,p)].
\end{eqnarray}
Two stream approximation has been confirmed to be valid by \citet{2003PhRvD..68f3001S,2012PhRvD..85f4004P}
and is much easier to deal with compared with the full Boltzmann equation,
so it is a prime choice for the calculation of
the energy density $u_{\nu,\bar\nu}$, the cooling rate $q_{\nu,\bar\nu}$ and the lepton number flux $F_{\nu,\bar\nu}$ of neutrinos and antineutrinos:
\begin{eqnarray}
u_{\nu,\bar\nu}=\frac{2\pi}{h^3} \int p^3(f_++f_-)_{\nu,\bar\nu}dp,\\
q_{\nu,\bar\nu}=\frac{1}{\sqrt{3}}\frac{2\pi c}{h^3}\frac{d}{dz} \int p^3(f_+-f_-)_{\nu,\bar\nu}dp,\\
F_{\nu,\bar\nu}=\pm\frac{1}{\sqrt{3}}\frac{2\pi c}{h^3}\int p^2(f_+-f_-)_{\nu,\bar\nu}dp.
\end{eqnarray}

\subsection{Numerical methods}
Now, we start to seek for a steady solution to the disk in which hydrostatic equilibrium, thermal balance and chemical balance are satisfied.
First, we assume an initial distribution of temperature $T^0(r,z)=8(r/r_{\rm in})^{1/3}$ MeV and election fraction $Y_{\rm e}^0(r,z)=0.4$
(In fact, our numerical calculation has shown that
the final convergent solution does not depend on the specific choice of the initial conditions
which only affects the speed of convergence of numerical calculation. Such independence proves the validity of our numerical methods in turn),
then we solve the vertical hydrostatic Eq.(8) subjected the boundary condition (7),
and it is not hard to obtain the zero order mass density distribution $\rho^0(r,z)$ and corresponding the thickness of the disk $H^0(r)$.
Then, we solve the two stream approximation Eq.(29) and (30), and obtain the neutrino/antineutrino spectra $f_{\pm(\nu,\bar\nu)}$.
With the neutrino/antineutrino spectra, we can solve the chemical evolution Eq.(22) and thermal evolution Eq.(24)
with the zero order mass density distribution $\rho^0(r,z)$ or the baryon number density $n_{\rm b}^0(r,z)$ fixed ,
until a steady state $\partial T^1/\partial t=0$ and $\partial Y^1_{\rm e}/\partial t=0$.
With the first order distribution $T^1(r,z)$ and $Y_{\rm e}^1(r,z)$,
we solve the vertical hydrostatic Eq.(8) subject to the boundary condition (7) once more
to get the corresponding first order mass density distribution $\rho^1(r,z)$.
Iterating this process until the final overall convergent solution of mass density $\rho(r,z)$, electron fraction $Y_e(r,z)$ and temperature $T(r,z)$ is obtained.

After that, we switch to the full Boltzmann equation Eq.(1) and (2) to solve the full energy dependent and
direction dependent distribution function of neutrinos and antineutrinos:
$[f_{+,-}(z,p,\mu)]_{,\nu}$ and $[f_{+,-}(z,p,\mu)]_{,\bar\nu}$.
With the full energy dependent and direction dependent spectra of neutrinos and antineutrinos,
we can calculate their annihilation luminosity precisely.
The annihilation rate of neutrino pairs is expressed as follows \citep{1997A&A...319..122R,2006NuPhA.777..356B}:
\begin{eqnarray}
Q(\nu_{\rm{e}}\overline{\nu}_{\rm{e}} )=\lefteqn{\frac{1}{4}\frac{\sigma_0c}{(m_{\rm{e}}c^2)^2(hc)^6}{}
\Bigg[ \frac{C_1+C_2}{3}\int_0^{\infty}dp\int_0^{\infty} dp'(p+p')(pp')^3
\int_{4\pi}d\Omega\int_{4\pi}d\Omega'f_{\nu_{\rm{e}}}f_{\overline{\nu}_{\rm{e}}}(1-\cos\Theta)^2}  \nonumber\\
&&{}+C_3(m_{\rm e}c^2)^2\int_0^{\infty}dp\int_0^{\infty}dp'(p+p')(pp')^2
\int_{4\pi}d\Omega\int_{4\pi}d\Omega'f_{\nu_{\rm{e}}}f_{\overline{\nu}_{\rm{e}}}(1-\cos\Theta)\Bigg].
\end{eqnarray}
where the typical cross section of neutrino interaction is $\sigma_0=1.705\times10^{-44}$ cm$^2$,
the weak interaction constants are $C_1+C_2\approx2.34$, $C_3\approx1.06$, $p$ and $p'$ is the energy of neutrinos and antineutrinos respectively,
$\Omega$ and $\Omega'$ is the solid angle of the incident direction of neutrinos and antineutrinos respectively,
$\Theta$ is the angle between neutrino beams and antineutrino beams (see Fig.1).
\begin{figure}
\begin{center}
\includegraphics[angle=270,width=0.5\textwidth]{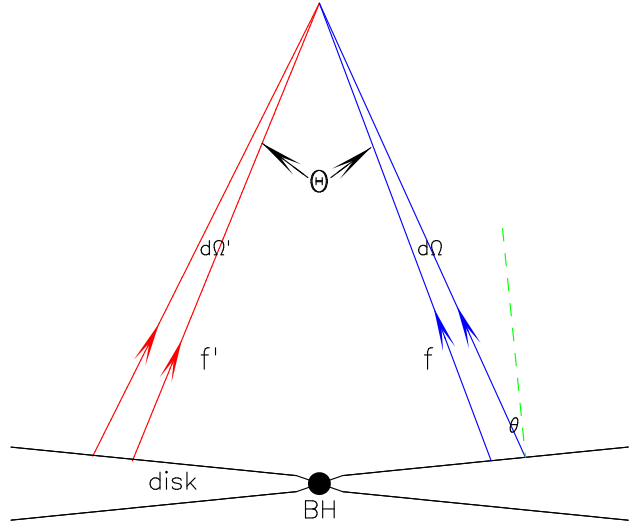}
\end{center}
\caption{The annihilation of neutrino pairs from the disk.}
\end{figure}

\section{Results}
In this section, taking the case $\dot M=10 M_{\rm sun}/s$, $\alpha=0.1$
as an example, we show the numerical results of the vertical structure of 
the accretion flow, its radial structure, the spectral energy distrition
of the neutrino/anti-neutrino from the disk and the final annihilation
luminosity of neutrinos and anti-neutrinos.

\subsection{The vertical structure of the disk}
\begin{figure}
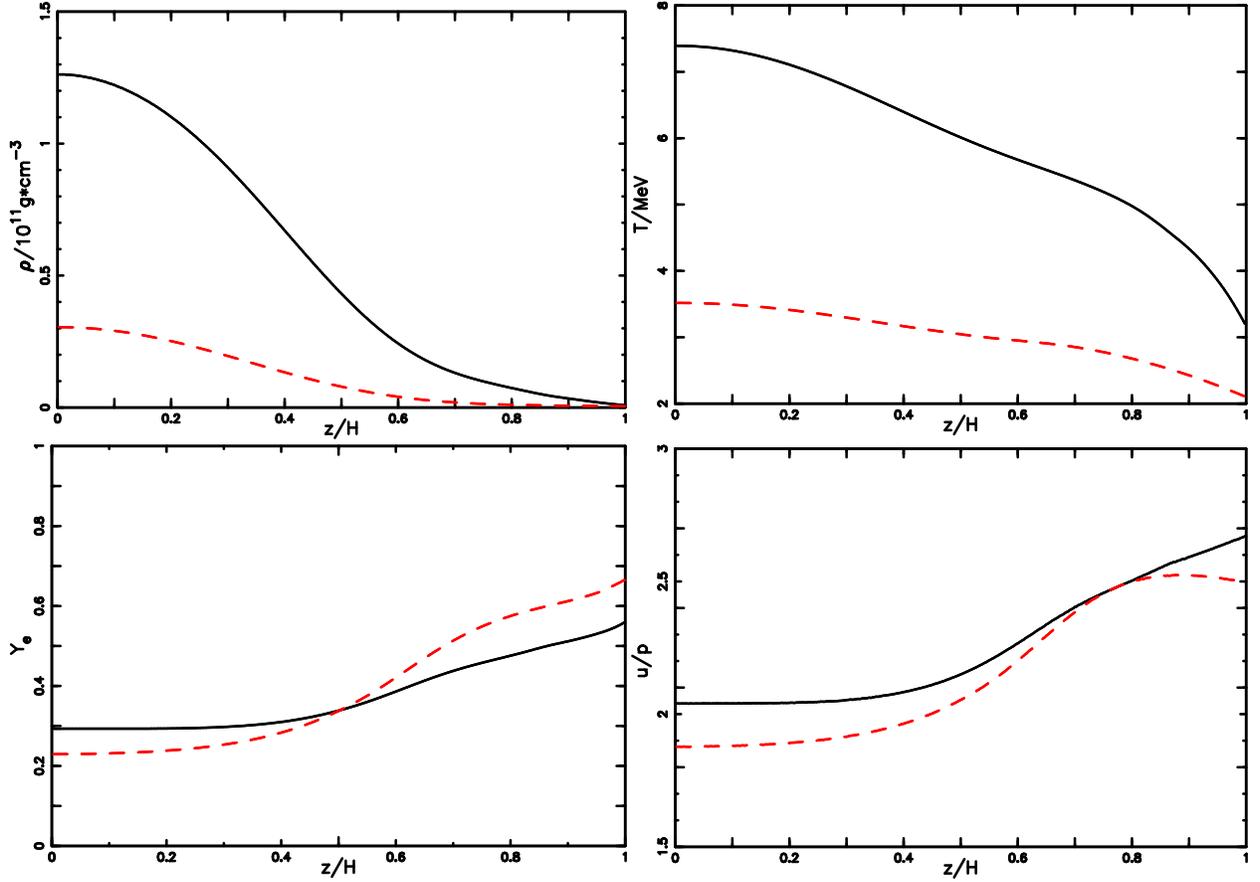

\includegraphics[angle=270,width=0.5\textwidth]{fig2a}%
\includegraphics[angle=270,width=0.5\textwidth]{fig2b}
\includegraphics[angle=270,width=0.5\textwidth]{fig2c}%
\includegraphics[angle=270,width=0.5\textwidth]{fig2d}
\caption{The vertical structure of the disk at different radius: $r=10 M$ (solid lines) and $r=35 M$ (dashed lines),
where $H$ is the thickness of the disk (see Fig.6)}
\end{figure}
Figure 2 shows the distribution of the mass density $\rho$, temperature $T$,
electron fraction $Y_e$
and the ratio of internal energy density to pressure $u/p$ in the vertical direction
at two different radius $10 M$ and $35 M$, respectively.

The mass density is about $\rho\sim10^{11}$ g/cm$^3$ and the temperature is about $T\sim8$ MeV in the inner part ($r=10 M$)
of the disk. The precise value of the temperature in the inner part of NDAFs is vitally important, which sensitively determines
the annihilation luminosity of neutrino pairs $L_{\nu\bar\nu}\sim T^9$ according to Eq.(34).

When the disk is in chemical equilibrium, the electron fraction cannot be describe by only a constant value throughout the disk, while
it varies a few times from the bottom to the surface of the disk.
It is the specific distribution of the electron fraction that guarantees the chemical equilibrium, which was not included in the most
previous works, as a result, the electron fraction has to be an artificial assumption there.

It is noticeable that the vertical distribution of $u/p$ and electron fraction $Y_{e}$ are positively correlated (Fig.2c and 2d).
The positive correlation implies that relativistic electrons dominate the pressure at the surface of the disk where $u/p\sim3$,
and non-relativistic protons and neutrons dominated the pressure at the bottom of the disk where $u/p\sim3/2$.
To justify the conclusion, we plot the vertical distribution of the ratio $p_i/p$ at radius $10 M$  and $35 M$ in the Fig.3,
where $p_1\equiv p_{\rm p}+p_{\rm n}$, $p_2\equiv p_{\rm e}+p_{\rm e^+}$ and $p_3\equiv p_{\rm rad}+p_{\nu}+p_{\bar\nu}$. It is indeed so, baryons (neutrons and protons) dominate the pressure at the
bottom of disk and leptons (electron pairs) dominated the pressure at the surface of the disk.
So it is not reasonable to assume a polytropic equation of state $p\propto \rho^{4/3}$ \citep{2010ApJ...709..851L}
which is the EOS of relativistic and strongly degenerate electron gas whose ratio of internal energy density to pressure is $u/p=3$.
If a polytropic equation of state is needed to do some approximation and estimation, actually $p\propto \rho^{5/3}$ is a better choice.
\begin{figure}
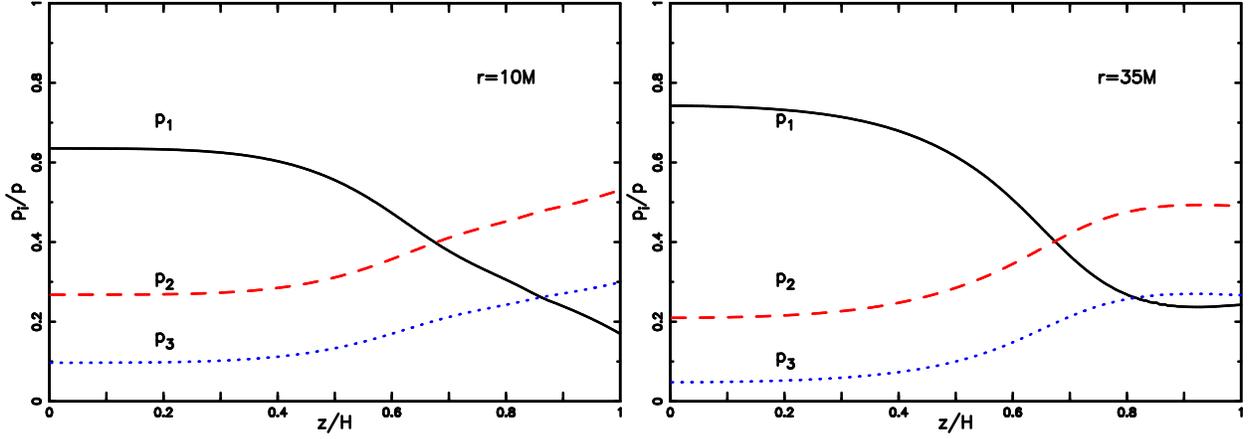

\includegraphics[angle=270,width=0.5\textwidth]{fig3a}%
\includegraphics[angle=270,width=0.5\textwidth]{fig3b}
\caption{The distribution of the percentage $p_i/p$ of different components: $p_1\equiv p_{\rm p}+p_{\rm n}$, $p_2\equiv p_{\rm e}+p_{\rm e^+}$ and $p_3\equiv p_{\rm rad}+p_{\nu}+p_{\bar\nu}$  in the vertical direction at different radius:
$r=10 M$ (left panel) and $r=35 M$ (right panel).}
\end{figure}

\subsection{The radial structure of the disk}
Two main assumptions are used in the above discussion:
we apply the self similar assumption of the distribution of mass density $\rho$ and internal energy density $u$ in the radial direction
when calculating the advection cooling term and we neglect the contribution of helium to the EOS
and the contribution of helium disintegration to the cooling term.
We now check their validity.
\subsubsection{Self-similar behavior in the radial direction}
Fig.4 shows the radial distribution of mass density $\rho$, internal energy density $u$, pressure $p$ and temperature $T$
on the equator plane $z=0$ and their corresponding fitting lines: $\rho\sim r^{-1.5}$,$u,p\sim r^{-2}$ and $T\sim r^{-0.6}$.
According to Fig.4, the self similar assumption of mass density $\rho$ and internal energy density $u$ in the radial direction is
rather a self consistent and accurate description.

Now we give a more physical explanation about the self-similar behavior in the radial direction: according to Fig.2 and Fig.3,
it is evident that non-relativistic protons and neutrons dominate total pressure at the bottom of the disk
and determine the surface density of the disk. Hence, we use the polytropic equation of state
$p\propto\rho^{5/3}$ to estimate vertical structure of the disk. Combining Eq.(7) and (8),
it is easy to get the scale thickness of the disk  $H\sim c_s/\Omega$, and the mass density $\rho\sim r^{-3/2}$.
In order to estimate the radial behavior of temperature and pressure, we must consider the more general EOS of non-relativistic gas  $p=\rho T$,
and the thermal balance between viscosity heating and neutrino cooling $q_+H\sim T^4$ or $p\Omega H\sim T^4$,
so $T^4\sim pc_s$, consequently $T\sim r^{-0.6}$ and $p\sim r^{-2.1}$ (see Fig4).

In order to gain an insight to the nature of the disk we investigate, we also calculate the advection factor
\begin{equation}
f_{\rm adv}\equiv\int q_{\rm adv} dz/\int q_+ dz,
\end{equation}
where $q_{\rm adv}$ and $q_+$ are the advection cooling rate and heating rate defined in the \S II, and the result is shown in Fig.5a. So it is evident that the disk is indeed a neutrino cooling dominated accretion disk as its name suggests: neutrino radiation dominates the cooling process and advection is always a minor role. The self similar assumption is only applied in the calculation of the advection term, so even if there is some small deviation between the realistic distribution and the self-similar assumption as shown in Fig.4, it has no much influence on the final results.
Thus the self similar assumption is rather a reasonable simplification.

\subsubsection{The fraction of He}
The number density of helium $n_{\rm He}$ is expressed as
\begin{equation}
n_{\rm He}=\frac{1}{4}(1-X_{\rm nuc})n_{\rm b},
\end{equation}
where the fraction of free nucleons $X_{\rm nuc}$ (protons and neutrons) is given by \citep{2007ApJ...664.1011J,1996ApJ...471..331Q}
\begin{equation}
X_{\rm nuc}=295.5\rho_{10}^{-3/4}T_{11}^{9/8}\exp{(-0.8209/T_{11})}.
\end{equation}
where $\rho_{10}$ is the mass density in unit of $10^{10}$ g/cm$^3$ and $T_{11}$ is the temperature in unit of $10^{11}$ K
and if $X_{\rm nuc}> 1$, then $X_{\rm nuc}= 1$.

We plot the free nucleons fraction of the gas on the equator plane $\log(X_{\rm nuc})$ calculated from Eq.(37) versus radius $\log (r/M)$ in Fig.5b.
According to Fig.5b, it is easy to know that $X_{\rm nuc}\gg1$, i.e., the assumption $X_{\rm nuc}= 1$ and $n_{\rm He}=0$ perfectly hold here.
So it is sound to neglect the contribution of helium to EOS and the contribution of the helium disintegration to disk cooling.

It should be noted that the radius where Helium dissociation becomes
important depends on the viscous parameter
$\alpha$ \citet{2007ApJ...657..383C}.
So in the case of smaller $\alpha$, the contribution of helium may not be
negligible.

\begin{figure}
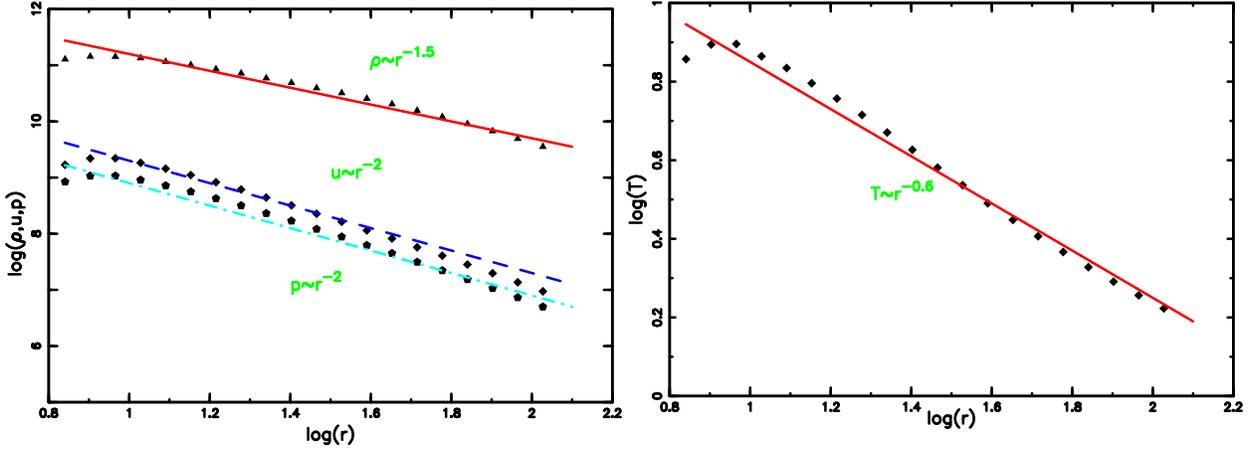

\includegraphics[angle=270,width=0.5\textwidth]{fig4a}%
\includegraphics[angle=270,width=0.5\textwidth]{fig4b}
\caption{\emph{Left panel}: The radial distribution of the mass density $\log (\rho/$g$\cdot$cm$^{-3})$, internal energy density $\log (u/$g$\cdot$cm$^{-3})$
and pressure $\log (p/$g$\cdot$cm$^{-3})$ versus $\log (r/M)$.
The straight lines are the corresponding linear fitting lines:
$\rho\sim r^{-1.5}$ (solid line), $u\sim r^{-2}$(dashed line) and $p\sim r^{-2}$ (dot-dashed line).
\emph{Right panel}: The radial behavior of the temperature $\log(T/\rm MeV)$ versus $\log(r/M)$
and the solid straight line is the corresponding fitting line $T\sim r^{-0.6}$.}
\end{figure}
\begin{figure}
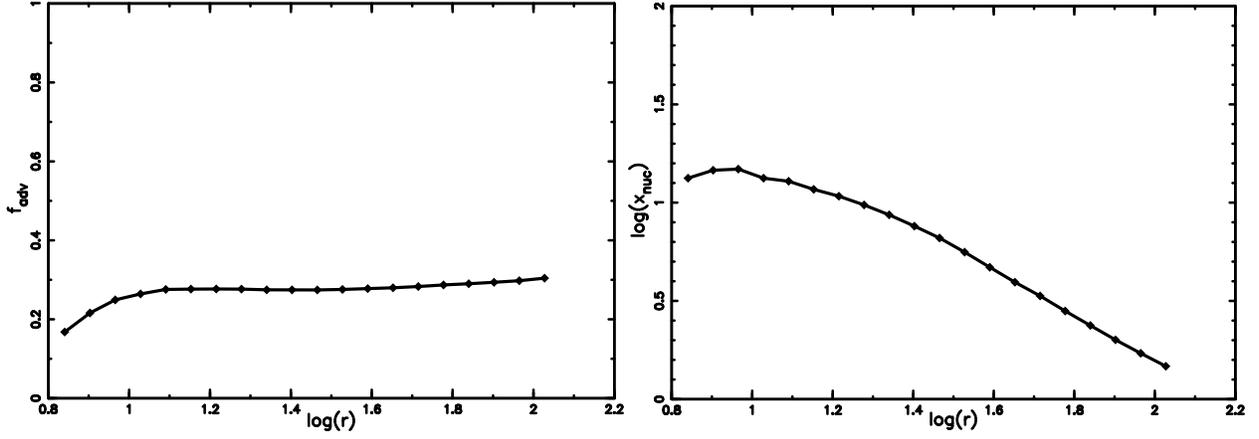

\includegraphics[angle=270,width=0.5\textwidth]{fig5a}%
\includegraphics[angle=270,width=0.5\textwidth]{fig5b}
\caption{(a)\emph{Left Panel}: The radial variation of the advection factor $f_{\rm adv}$;
(b)\emph{Right Panel}: The radial distribution of free nucleons fraction $\log(X_{\rm nuc})$ calculated from Eq.(37).}
\end{figure}

\subsubsection{Profile of the disk}
In addition, we plot the radial distribution of the surface density $\sigma$ in Fig.6a and the profile of the disk thickness $H(r)$ in Fig.6b.
Here, the thickness $H(r)$ is defined according to the mass density contrast $\rho(z=H)=\rho(z=0)/100$.
$H(r)$ serves as the upper boundary of Boltzman equation (see \S2.1),
and it is different from definition of the usual scale thickness of the
disk which is mostly used in the one-dimensional disk model.
According to Fig.6, the disk we consider is a geometrically thick disk with thickness $H\approx r$, which will contribute to higher annihilation efficiency as we will discuss
in the next section.
\begin{figure}
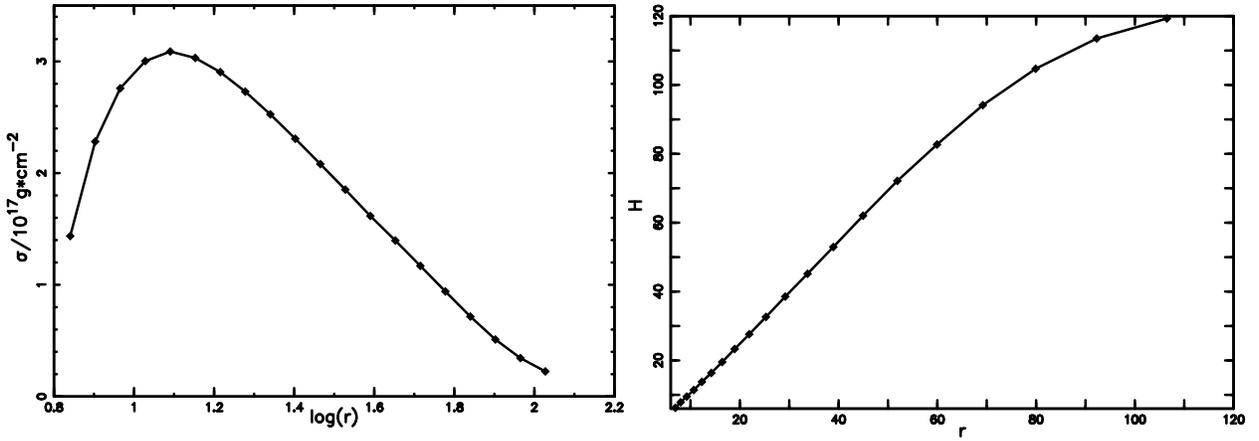

\includegraphics[angle=270,width=0.5\textwidth]{fig6a}%
\includegraphics[angle=270,width=0.5\textwidth]{fig6b}
\caption{(a)\emph{Left panel}:The radial distribution of surface density $\sigma$ in unit of $10^{17}$ g/cm$^2$;
(b)\emph{Right panel}: The profile of the disk thickness $H(r)$ versus radius $r$.}
\end{figure}

\subsection{Neutrino spectrum}
\begin{figure}
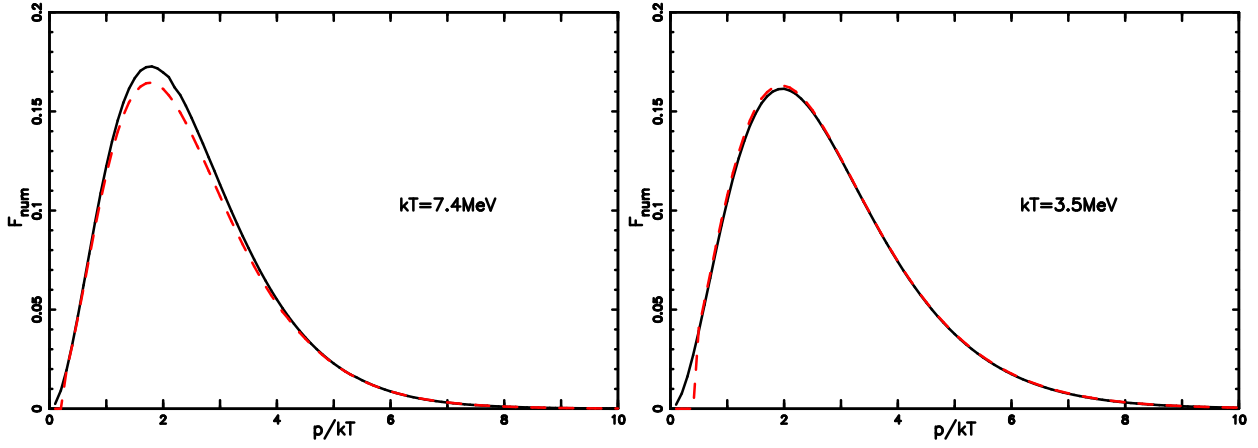

\includegraphics[angle=270,width=0.5\textwidth]{fig7a}%
\includegraphics[angle=270,width=0.5\textwidth]{fig7b}
\caption{The direction-averaged spectra of neutrino (solid lines) and antineutrino (dashed lines)
at different radius:$r=10 M$ (left panel) and $r=35 M$ (right panel).}
\end{figure}
When the disk is in chemical equilibrium, the flux of lepton number $F_{\rm lep}=F_{\nu}+F_{\bar\nu}$ vanishes, i.e.:
\begin{equation}
\int p^2(f_+-f_-),_{\nu}\mu dpd\mu=\int p^2(f_+-f_-),_{\bar\nu}\mu dpd\mu.
\end{equation}
Figure.7 shows the direction averaged neutrino/antineutrino spectrum
\begin{equation}
F_{\rm num}=\left(\frac{p}{kT}\right)^2\int f\mu d\mu.
\end{equation}
at the surface of the disk at different radius, where $f=f_+(H,p,\mu)$, $T=7.4$ MeV for $r=10M$ and $T=3.5$ MeV for $r=35M$ (see Fig.2b).
Then, at the surface of the disk the chemical equilibrium condition Eq.(38) is transformed to be
\begin{equation}
\int F_{\rm num},_{\nu}dp=\int F_{\rm num},_{\bar\nu}dp.
\end{equation}
So, chemical equilibrium only requires that the total amount of neutrino number flux and antineutrino number flux are equal,
i.e. integration of $F_{\rm num}$ of neutrinos and neutrinos over energy $p$ are equal,
while it is interesting that the form $F_{\rm num}$ of neutrinos and antineutrinos almost coincide with each other in fact.

In addition, it is necessary to explain that there is a cusp on the antineutrino spectrum in the lower energy end: it is easy to
know that there is a lower energy limit $E_{\rm min}=m_{e^+}+m_n-m_p$ for antineutrinos from the Urca process
$\overline{\nu}_e + p \leftrightarrow e^+ + n$ which gives rise to the cusp on the energy spectrum,
while there is no such an energy limit for neutrinos from the process $\nu_e+n\leftrightarrow p+e$,
so the neutrino energy spectrum is smooth as expected.

\subsection{Annihilation luminosity}
In order to calculate the neutrino/antineutrino luminosity and their corresponding luminosity, we have to
make a discount on the vertical structure of the disk: we assume the disk to be lying on the equator plane,
and then modify the resulting annihilation luminosity by taking the thickness of the disk into consideration.
With the thin disk simplification, all the elements in the Eq.(34) for the annihilation rate are available (see Fig.1), so it is easy to
numerically do the integration of Eq.(34) over the entire surface of the disk and the whole energy span of neutrinos and antineutrinos.

In addition, the neutrino/antineutrino energy flux $\tilde F_{\nu,\bar\nu}$, luminosity at the surface of the disk $L_{\nu},L_{\bar\nu}$
and their annihilation luminosity $L_{\nu\bar\nu}$
are expressed as
\begin{eqnarray}
\tilde F_{\nu,\bar\nu}=\frac{2\pi c}{h^3}\int\!\int p^3f_{\nu,\bar\nu}\mu dpd\mu,\\
L_{\nu}=2\int_{r_{\rm in}}^{r_{\rm out}}2\pi r \tilde F_{\nu} dr,\\
L_{\bar\nu}=2\int_{r_{\rm in}}^{r_{\rm out}}2\pi r \tilde F_{\bar\nu} dr,\\
L_{\nu\bar\nu}=2\int_{H(r)}^{\infty}\!\int_{0}^{\infty} 2\pi r Q_{\nu\bar\nu}  dz dr.
\end{eqnarray}
where $r_{\rm in}=6 M$, $r_{\rm out}=100 M$ in our calculation and $H(r)$ is the thickness of the disk at radius $r$ (see Fig.6b).

The results are as follows:
$L_{\nu}\approx L_{\bar\nu}=5.2\times10^{53}$ ergs/s, $L_{\nu\bar\nu}=1.66\times10^{51}$ ergs/s,
the annihilation efficiency $\eta\equiv L_{\nu\bar\nu}/(L_{\nu}+L_{\bar\nu})=0.16\%$.
While it is not the final result, we must take into consideration of the thickness of disk $H\approx r$ (see Fig.6b)
to correct the thin disk simplification.

Fig.8 shows the concrete distribution of the annihilation rate $Q_{\nu\bar\nu}$ calculated with the thin disk simplification in the $r,z$ plane.
The distribution of the annihilation rate is nearly isotropic aside the directions occupied by the disk
and the half open angle of empty funnel along the central axis of the disk is about $45$ degrees which is determined by the ratio of thickness
to radius of the disk $H/r\approx 1$. The major contribution to the total annihilation luminosity is concentrated at the zone near the central black hole
of the disk. The solid angle that the assumed thin disk lying on the equator plane subtends to the zone is about $\Omega_{\rm thin}\approx2\pi$,
while solid angle that the realistic thick disk subtends to that zone is about $\Omega_{\rm thick}\approx2\pi(1+\cos(45^{\circ}))\approx1.7\Omega$,
and so final annihilation luminosity should be multiplied by a modification factor $(\Omega_{\rm thick}/\Omega_{\rm thin})^2\approx3$.

Hence, the final results are as follows:
neutrino/antineutrino luminosity is about $L_{\nu}=L_{\bar\nu}=5.2\times10^{53}$ ergs/s,
annihilation luminosity is about $L_{\nu\bar\nu}=5\times10^{51}$ ergs/s
and annihilation efficiency is about $\eta=0.48\%$.

\begin{figure}
\begin{center}
\includegraphics[width=0.5\textwidth]{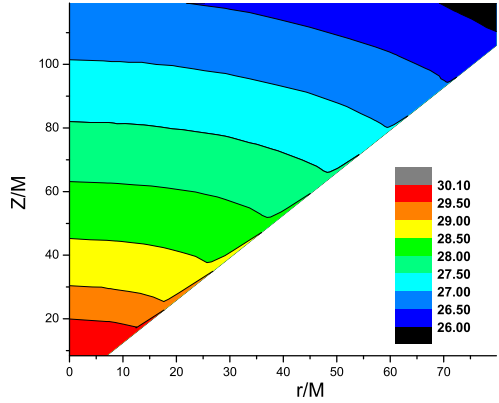}
\end{center}
\caption{Contour of annihilation rate $\log Q_{\nu\bar\nu}$/(ergs$\cdot$ cm$^{-3}\cdot$s$^{-1}$) in $r,z$ plane, and the blank zones are occupied by the disk.}
\end{figure}

\section{Conclusions and Discussions}
Temperature of the inner part of NDAFs sensitively determines the final luminosity
of neutrino annihilation as $L_{\nu\bar\nu}\propto T^9$.
In the extreme model of this paper ($M=3.3M_{\rm sun}$, $\dot M=10 M_{\rm sun/s}$ and $\alpha=0.1$),
it is found that temperature in the inner part of the disk is about $T\approx8$ MeV,
and the corresponding annihilation luminosity is about $L_{\nu\bar\nu}\approx5\times10^{51}$ ergs/s
which roughly satisfies the energy demand of the most energetic GRBs.
If the temperature of disks changes from 8 MeV to 4MeV, the annihilation luminosity reduces
about 500 times, which is not enough to power the fireball of the most
energetic  GRBs with the isotropic luminosity about $10^{52}$ ergs/s.
In the standard model of GRBs, the duration of the energy
injection is the duration of the GRB prompt emission, which could be about 2s for
short GRBs and about 100-1000s for long bursts. Assuming an extreme constant accretion rate 
of 10 Msun/s means therefore an enormous total mass consumption for powering GRBs,
especially for powering long bursts, which might be realistic. Therefore,
the simple NDAF model which is investigated in
this work can not produce sufficient energy to power GRBs,
the effects of the spin of black hole or/and the magnetic field in the
accretion flow might be introduced to power the central engine of GRBs
\citep[see][for instance]{2002ApJ...567..463L,2007ApJ...657..383C,2009ApJ...700.1970L,2010ScChG..53S..98L,2010A&A...509A..55J}.

It is very easy to understand that the temperature of the inner disk mainly depends on the
accretion rate of the flow. Besides the extreme model, we also investigate the models with
the lower accretion rate, such as  $\dot M =0.1 M_{\rm sun}$/s and $1 M_{\rm sun}$/s,
and the results including the resulting temperature $T$ in the inner part of disk $(r=10M)$,
neutrino and antineutrino luminosity $L_{\nu}+L_{\bar \nu}$, annihilation luminosity $L_{\nu\bar\nu}$
and the corresponding annihilation efficiency $\eta$ are listed in Table 1.
For all the three different accretion rate $10, 1, 0.1 M_{\rm sun}$/s, the thickness $H(r)$ 
we defined in \S4.2.3 is roughly equal to radius $r$.
\begin{table}
\caption{Luminosity of neutrino annihilation with different accretion rates}
\begin{tabular}{cccccc}
\hline
\hline
$\dot M(M_{\rm sun}/{\rm s})$ & $T_{r=10M}(\rm MeV)$ & $L_{\nu}+L_{\bar \nu}(\rm ergs/s)$ & $L_{\nu\bar\nu}(\rm ergs/s)$ &
$\eta\equiv L_{\nu\bar\nu}/(L_{\nu}+L_{\bar \nu})$\\
\hline
10.&7.4&$1.04\times10^{54}$&$5.0\times10^{51}$&$4.8\times10^{-3}$\\
1.0&4.2&$0.92\times10^{53}$&$3.3\times10^{49}$&$3.6\times10^{-4}$\\
0.1&3.3&$0.30\times10^{52}$&$4.0\times10^{46}$&$1.3\times10^{-6}$\\
\hline
\end{tabular}
\end{table}

It is obvious that, accretion rate sensitively determines the annihilation luminosity of neutrino pairs and
the annihilation luminosity will not exceed $10^{50}$ ergs/s when the corresponding accretion rate is lower
than $1M_{\rm sun}$/s.
So NDAFs with accretion rate lower than $1M_{\rm sun}$/s are unlikely to serve as central engines of GRBs.

Neutrino and antineutrino spectra is the second major factor determining the annihilation luminosity.
The resulting neutrinos and antineutrino spectra obtained by solving the Boltzmann equation shows that,
when the disk is in chemical equilibrium, the emission of neutrinos and antineutrinos are almost symmetric
with nearly identical energy spectra,
but the spectra is neither in the form of black body nor in the form of the gray body.
And as shown in \cite{2012PhRvD..85f4004P}, the black body spectra of neutrino
and the neutrino spectra based on the most commonly used simplified model of neutrino
transport \citep{2002ApJ...579..706D}
or equivalently the gray body spectra \citep{2007ApJ...664.1011J}
can overestimate the annihilation luminosity by nearly one order of magnitude.
In the following, we will also check the validity of the previous assumption on
the neutrino transport.

It is easy to estimate the mean optical depth of neutrinos in the inner part of the disk.
From Fig. 6a and Fig. 2b, at $r=10 M$,
the surface density is $\sigma=3\times10^{17}$ g/cm$^2$ and the temperature is $T=7.4$MeV.
According to \cite{2002ApJ...579..706D}, the neutrino opacity of absorption $\tau_a$ and scattering $\tau_s$ are expressed as
\begin{equation}
\tau_s=2.7T_{11}^2\sigma_{17},
\end{equation}
\begin{equation}
\tau_a=4.5T_{11}^2\sigma_{17},
\end{equation}
where $T_{11}$ is the temperature in unit of $10^{11}$K and $\sigma_{17}$ is the surface density in unit of $10^{17}$ g/cm$^2$.
Therefore, the optical depth of neutrinos we get is
$\tau_a=10,\tau_s=6$, so the disk is optically thick for neutrinos at $r=10M$.
However, as \cite{2012PhRvD..85f4004P} has shown in quasi-optically opaque case ($\tau_{\rm a,s}=0.1\sim1$),
the neutrino spectra are neither the black body spectra $f_{\rm black}$ \citep{1999ApJ...518..356P}
nor the gray body spectra $f_{\rm gray}$ \citep{2002ApJ...579..706D,2007ApJ...664.1011J}:
\begin{eqnarray}
f_{\rm black}&=&\frac{1}{\exp{(p/kT)}+1} ,\\
f_{\rm gray}&=&\frac{b}{\exp{(p/kT)}+1} ,
\end{eqnarray}
and the block factor $b$ is given by
\begin{eqnarray}
b=\frac{1}{(3/4)(\tau/2+1/\sqrt{3}+1/3\tau_a)},
\end{eqnarray}
where $\tau=\tau_a+\tau_s$, so here $b=0.16$.

For comparison, we plot the direction-averaged spectra $F_{\rm num}$ of the black body spectra,
the gray body spectra and the more realistic spectra obtained by solving Boltzmann equation (see Fig.9).
According to Fig.9, it is obvious that the first assumption on neutrino transport
overestimates the neutrino luminosity about $30\%$
%(estimation by comparing the peak height of their spectra)
and the corresponding annihilation luminosity about $(1.3^2-1)\approx70\%$,
while the second assumption underestimates the neutrino luminosity about 5 times and the corresponding
annihilation luminosity about 25 times.
It may explain why \citet{2002ApJ...579..706D} drew a conclusion that the annihilation luminosity of NDAFs is no more than $10^{50}$ ergs/s, so NDAFs cannot serve as the central engine of GRBs, on the contrary, \citet{1999ApJ...518..356P} claimed an opposite conclusion.

\begin{figure}
\begin{center}
\includegraphics[angle=270,width=0.4\textwidth]{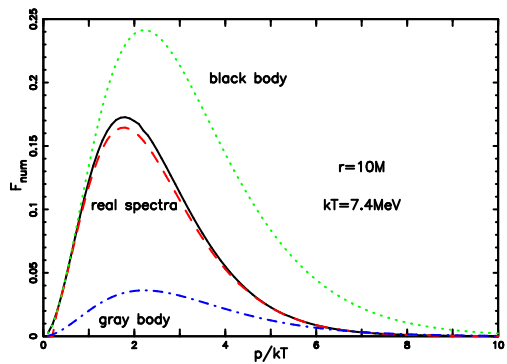}
\end{center}
\caption{Comparison of the direction averaged black body spectra, gray body spectra and the real spectra obtained by solving the Boltzmann equation.}
\end{figure}

The thickness of the disk is the third factor that affects the final annihilation luminosity: a larger ratio of thickness to radius means a larger
solid angle the disk subtends to the annihilation zone. In the case we consider,
the thick disk ($H\approx r$) will enhance the annihilation luminosity about 3 times than that of a thin disk.

The significance of the distribution of electron fraction in NDAFs has been discussed by many previous works \citep{2002ApJ...577..311K,2005ApJ...632..421L,2007ApJ...661.1025L,2005ApJ...629..341K}.
In this work, the distribution of electron fraction is a natural result of chemical equilibrium, thermal balance and hydrostatic equilibrium,
rather than an artificial assumption in most previous works, so which should be most reliable.

As shown in Fig. 5, helium is completely absent within $100 M$ of NDAFs, so it is not needed
to consider the contribution of helium to the total pressure and internal energy
or the contribution of helium disintegration to the cooling of the disk.

It should be emphasized that our calculation also has its own limitations.
First, in this work, the disk is thick and has two dimensional structure,
while the neutrino transport is treated in one dimension.
Second, the effects of the motion of accretion flow and the curved spacetime
on the vertical transport of neutrinos are neglected.
These effects are significant in the inner part of the disk, for example,
at $r=10M$, the special relativity correction to the neutrino energy can be $v/c\sim30\%$,
and the general relativity correction to the energy can be $M/r\sim10\%$.
Third, in the zone near to the event horizon of the central black hole where the annihilation concentrates,
the trajectories of neutrinos are severely bent by the central black hole and a considerable amount
of neutrinos will undoubtedly be captured by the hole.
In addition, the gravitational instability of the disk
was proposed by \citep{2006ApJ...636L..29P} to explain the energetic X-ray flares after the prompt emission in GRBs.
The outer region of NDAFs with extremely high accretion rate is
gravitational unstable \citep{2007ApJ...657..383C}, will fragment and cause a variable accretion rate in the inner region.
%Then the axially symmetry and steady accretion assumption should be modified.

%Even the electron neutrino oscillation becomes an non-negligible element at this length scale and will suppress the annihilation luminosity by a few times.
%Considering all these limitations together, the results listing in the above table is only an estimation of order of magnitude.
%So two dimensional disk structure, two dimensional neutrino transport, relativistic correction to the neutrino and antineutrino spectrum
%and their trajectories, and the neutrino oscillation are indispensable for the precise annihilation luminosity,
%while, all of them are still untouched fields.

\acknowledgements
We would like to thank the annonymous referee for her/his constructive
suggestions and comments which are helpful for the improvement of
this paper.
This work is partially supported by
National Basic Research Program of China (2009CB824800, 2012CB821800),
the National Natural Science Foundation (11073020, 11133005),
and the Fundamental Research Funds for the Central Universities (WK2030220004).

\end{document}